# Search for Heavy Majorana Neutrinos at LHC using Monte Carlo simulation

**H.M.M.Mansour[1,2], Nady Bakhet[1,3]**

(Affiliation): [1]Department of Physics, Faculty of Science, Cairo University
Email: [2]mansourhesham@yahoo.com, [3]nady.bakhet@cern.ch, nady.bakhet@ yahoo.com

**Abstract**
Heavy neutrinos can be discovered at LHC. Many extensions for Standard Model predict the existence of a new neutrino which has a mass at high energies. B-L model is one of them which predict the existence of three heavy (right-handed) neutrinos one per generation, new gauge massive boson and a new scalar Higgs boson which is different from the SM Higgs. In the present work we search for heavy neutrino in 4 leptons + missing energy final state events which are produced in proton-proton collisions at LHC using data produced from Monte Carlo simulation using B-L model at different center of mass energies. We predict that the heavy neutrinos pairs can be produced from $Z'_{B-L}$ new gauge neutral massive boson decay and then the heavy neutrino pairs can decay to 4 leptons + missing energy final state which give us an indication for new signature of new physics beyond Standard Model at higher energies at LHC.

**Keywords:**

## I. INTRODUCTION

Standard Model of elementary particles needs an extension to explain some facts. One of these facts is that the neutrinos are massive. $B-L$ model (baryon number minus lepton number) [1] is a simple extension of the Standard Model which plays an important role in various physics scenarios beyond the Standard Model (SM). Firstly, the gauged $U(1)_{B-L}$ symmetry group is contained in a Grand Unified Theory (GUT) described by a SO(10) group. Secondly, the scale of the $B-L$ symmetry breaking is related to the mass scale of the heavy right-handed Majorana neutrino mass [2-3] terms providing the well-known see-saw mechanism of light neutrino mass generation. Three heavy singlet (right-handed) neutrinos $v_h$ are invoked. $B-L$ symmetry breaking has been considered, based on the gauge group. The model under consideration $B-L$ symmetry-breaking is based on the gauge group

$G_{B-L} = SU(3)_C \times SU(2)_L \times U(1)_Y \times U(1)_{B-L}$

This model provides a natural explanation for the presence of three right-handed neutrinos [4] and can account for the current experimental results of the light neutrino masses and their mixings. The heavy neutrinos are rather long-lived particles producing distinctive displaced vertices that can be seen in the detectors. Lastly, the simultaneous measurement of both the heavy neutrino mass and decay length enables an estimate of the absolute mass of the parent light neutrino. An attractive feature of the $B-L$ model is that a successful realization of the mechanism of baryogenesis through leptogenesis to explain the observed matter-antimatter asymmetry depends on two main parameters, the mass of $Z'_{B-L}$ and the coupling constant $g'_1$. Therefore, the model B-L is controlled by two parameters:

| ψ | $SU(3)_C$ | $SU(2)_L$ | Y | B-L |
|---|---|---|---|---|
| $q_L$ | 3 | 2 | 1/6 | 1/3 |
| $u_R$ | 3 | 1 | 2/3 | 1/3 |
| $d_R$ | 3 | 1 | -1/3 | 1/3 |
| $l_L$ | 1 | 2 | -1/2 | -1 |
| $e_R$ | 1 | 1 | -1 | -1 |
| $V_R$ | 1 | 1 | 0 | -1 |
| H | 1 | 2 | 1/2 | 0 |
| χ | 1 | 1 | 0 | 2 |

**TABEL I.** Y and B−L quantum number assignation to chiral fermions and scalar fields.

the mass of the $Z'_{B-L}$ and the coupling constant $g'_1$ determining $Z'_{B-L}$ couplings [5-6]. Two experimental constraints exist on these two parameters. The first comes from direct search for heavy neutral gauge bosons at the Fermi Lab. which excludes a $Z'_{B-L}$ mass less than 600 GeV. The second limit comes from LEP

$$\frac{M_{Z'_{B-L}}}{g'_1} \geq 6 \text{ Tev} \qquad (1)$$

This constraint will provide an upper bound to the $Z'_{B-L}$ production cross sections at the LHC. Table(1) gives the quantum numbers in the B-L model.

For the heavy neutrino masses $M_{v_h}$ we will take them to be degenerate and relatively light, varying in the range 50 GeV < $M_{v_h}$ < 500 GeV. For this analysis of the heavy neutrino of B − L model we have chosen different values of $Z'_{B-L}$ masses and $g'_1$ = 0.2 and different LHC center of mass energies.

## II. RESULTS
### A. Branching Ratios of Heavy neutrinos

The heavy neutrino has a small mixing with the light neutrino and so it provide very small couplings to gauge and Higgs bosons which gives all the decay channels of $V_h$:

$$v_h \to W^\pm + l^\pm$$
$$v_h \to z + v$$
$$v_h \to H_1 + v$$
$$v_h \to H_2 + v$$

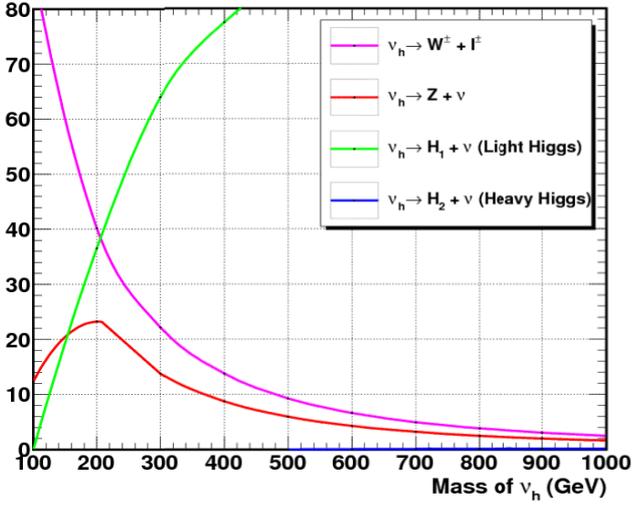

**FIG. 1.** Heavy neutrino branching ratios versus its mass at $M_{H_1}$ = 120 GeV and $M_{H_1}$ = 450 GeV and $g'_1$ = 0.2 .

Fig. 1 presents all branching ratios for the different decay channels versus different heavy neutrino masses. B-L model includes $H_1$ SM-like Higgs boson and $H_2$ heavy Higgs boson. One can notice from this fig. that the ratio of the decay channel $BR(v_h \to W^\pm + l^\pm)$ is the highest ratio in comparison with the other decay channels. Also, the ratio of the decay channel $BR(v_h \to H_1 + v)$ increase with increasing the heavy neutrino mass and the decay channel $v_h \to z + v$ has a high percent level at heavy neutrino mass of 200GeV. The ratio of $v_h \to H_2 + v$ decay channel is very small and we can neglect it.

### B. Production cross sections

In this section we determine the production cross section of heavy neutrino discovery at LHC for various CM energies, 5,7,10 and 14 TeV using MadGraph5/ MadEvent and PYTHIA8 programs [7-9]. We consider the production channel of heavy neutrino pair production via the $Z'_{B-L}$ boson decay. The distinctive features of the B - L model take place because the heavy neutrinos decay predominantly to SM gauge bosons in association with a lepton (either charged or neutral, depending on the electrical nature of the SM gauge boson). Also, once heavy neutrinos are pair-produced via the $Z'_{B-L}$ boson, they give rise to novel and spectacular multi-lepton decay modes of the intermediate boson. Thus the rate for the pair production of the heavy neutrinos depends on the mass of the $Z'_{B-L}$ and the strength of the B - L coupling $g'_1$.

The pair production of heavy neutrino via $Z'_{B-L}$ gauge boson exchange in the s-channel has the following Feynman diagram

$$pp \to Z'_{B-L} \to v_h v_h$$

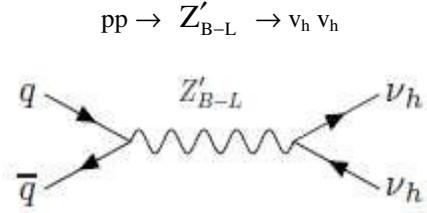

**FIG. 2.** Feynman diagram for Heavy Neutrino pair production.0

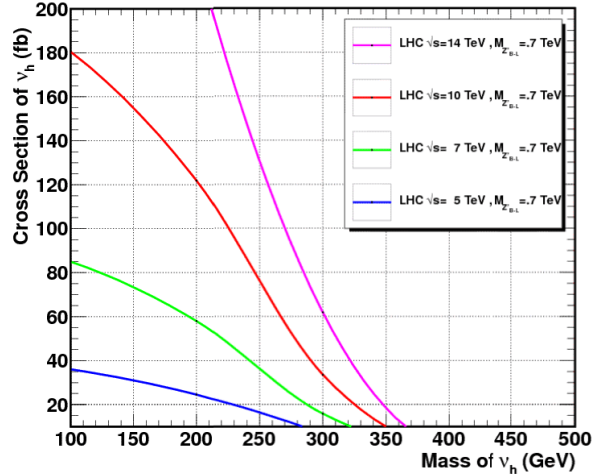

**FIG. 3.** Heavy neutrino pair production cross sections s at the LHC (for √s = 5,7,10 and 14 TeV, $Z'_{B-L}$ mass = 0.7 TeV and $g'_1$ = 0.2).

The process $pp \to Z'_{B-L} \to v_h v_h$ can be tested d at the LHC (for √s =5,7,10 and d 14 TeV CM energy) if we take the value of the total integrated luminosity 1 fb-1 and take the maximum value of the cross section to be 180 fb for mass of $Z'_{B-L}$ = 700 GeV and $g'_1$ = 0.2. For mass of $v_h$ = 100 GeV as in fig. 3 we can say that approx. 180 events are produced according to the relation

$$N = L^\sigma \qquad (2)$$

Where N is the number of events and L is the total integrated luminosity and s is the production cross section. Also we can deduce from fig. 3 that the production section depends on the $Z'_{B-L}$ boson mass and the value of the $g'_1$ coupling and as we will see later in addition to the LHC CM energy.

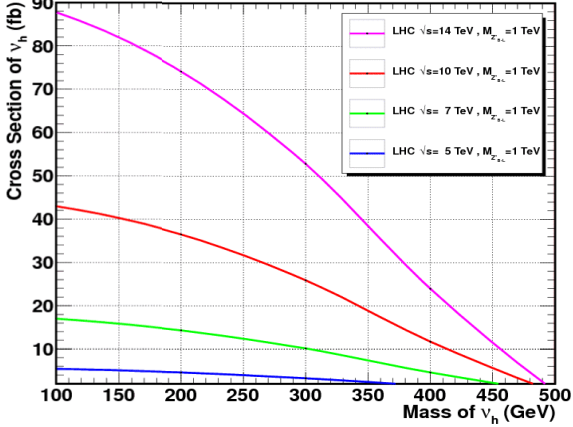

**FIG. 4.** Heavy neutrino pair production cross sections s at the LHC (for √s =5,7,10 and 14 TeV, $Z'_{B-L}$ mass = 1 TeV and $g'_1$ = 0.2).

Fig. 4 shows that When $Z'_{B-L}$ mass is 1 TeV and LHC CM energy is 10 TeV and $g'_1$ = 0.2 the e maximum production cross section is 45 fb . Comparing this by fig. 3 we find that the cross section is 180 fb at $Z'_{B-L}$ mass 0.7 TeV and $g'_1$ = 0.2,hence the cross section of the heavy neutrino pair production depend on $Z'_{B-L}$ mass and decrease with the increase of $Z'_{B-L}$ mass.

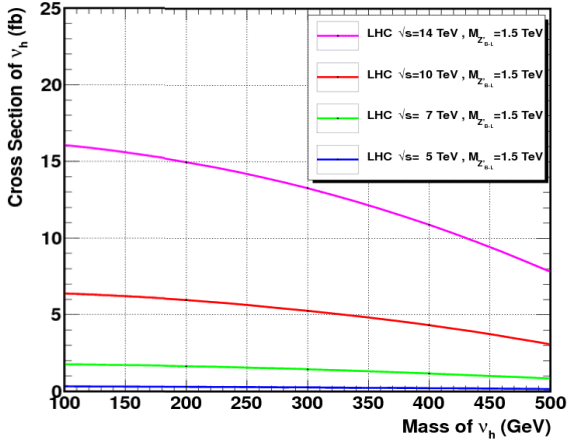

**FIG. 5.** Heavy neutrino pair production cross sections s at the LHC (for √s = 5, 7, 10 and 14 TeV, $Z'_{B-L}$ mass = 1.5 TeV and $g'_1$ = 0.2).

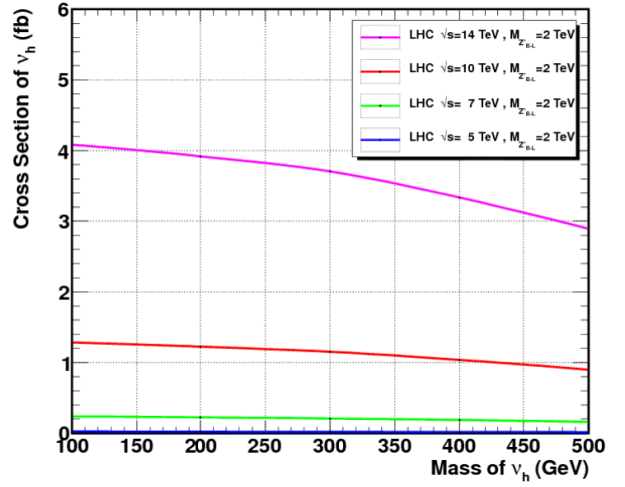

**FIG. 6.** Heavy neutrino pair production cross sections s at the LHC (for √s =5,7,10 and 14 TeV, $Z'_{B-L}$ mass = 2 TeV and $g'_1$ = 0.2).

Similar results are shown in fig's 5 and 6.From figs. 3-6 we note that the production cross section decrease with increasing $Z'_{B-L}$ mass at fixed value of $g'_1$. So the production cross section is enhanced due to the resonant contribution from the $Z'_{B-L}$ exchange in n the s-channel but falls rapidly with increasing heavy neutrino mass. Fig. 7 shows the behavior of the cross section for different values of the coupling constant and energies.

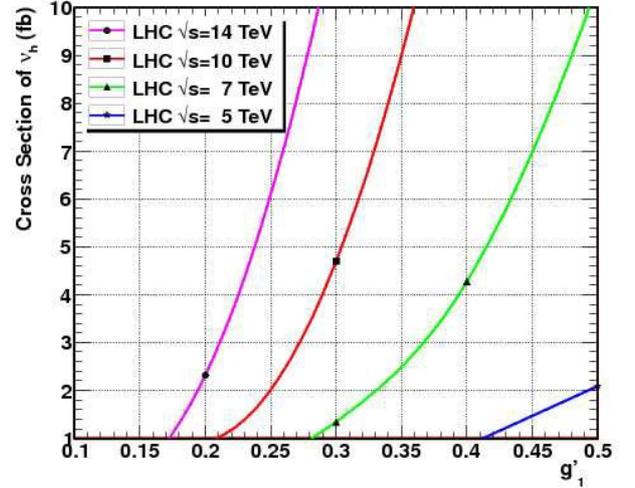

**FIG. 7.** Heavy neutrino pair production cross sections at the LHC for different values of $g'_1$ (for √s =5,7,10 and 14 TeV, $Z'_{B-L}$ mass = 1.5 TeV).

The dominant production mode for the heavy neutrinos at the LHC will be through the Drell-Yan mechanism [10] with $Z'_{B-L}$ in the s-channel.

**C. Heavy neutrino decays**

The detection of the signal coming from the pair production of the heavy neutrinos in the B-L model at LHC will be through very clean signal for four charged leptons in the final states and missing energy due to the associated neutrinos. The first heavy neutrino dominant decay is to a $W^+$ boson and a charged lepton here is $e^+$ and the second heavy neutrino decays to Z boson and a light neutrino. We can have the following final states as our signal (see fig.8 below)

$$pp \to Z'_{B-L} \to \nu_h \nu_h \to \mu^- \mu^+ e^- e^+ + \not{E} \quad (3)$$

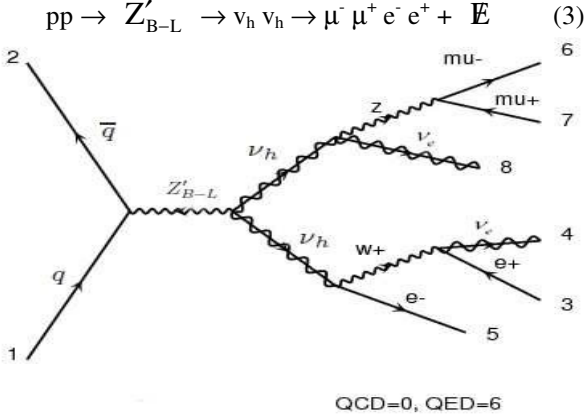

**FIG. 8.** Feynman diagram for heavy neutrino pair production and its decay in B-L model.

In our simulation for heavy neutrino detection at LHC using MadGraph5/Madevent and PYTHIA8 we will put some kinematics cuts to be taken into account when selecting the final states with four leptons and missing energy (momentum):

(a) For the charged leptons momentum and pseudo rapidity we take:
$$p_T^l > 20 \text{ GeV and } |\eta| < 2.5$$

(b) The minimum missing transverse energy cut is
$$\not{E} > 20 \text{ GeV}$$

(c) For resolving different leptons in the detector we put
$$\Delta R_{l_i l_j} \geq 0.2$$

(d) The minimum cut for the invariant mass for lepton pair will be
$$M_{l_i^+ l_i^-} > 10 \text{ GeV}$$

And for our simulation we choose the total integrated Luminosity:
$$\int L dt = 1 \text{fb}^{-1}$$

We will plot the various kinematics distributions for the signal arising from the process satisfying the above selection cuts for two different values of heavy neutrino masses at LHC $\sqrt{s}$ = 14 TeV CM energy $Z'_{B-L}$ mass = 1.5 TeV and $g'_1$ = 0.2

Also to simulate our process to produce final state with four charged leptons and missing energy at LHC by using PYTHIA8 we use the following commands after generating matrix elements for this process by MadGraph5/Madevent generator:

PartonLevel: all = On
PartonLevel: MPI = On    Open multiparton interactions
PartonLevel: ISR = On    Open Initial State Radiation
PartonLevel: FSR = On    Open Final State Radiation
HadronLevel:Hadronize = On   Open for hadronization

Turn off all decay channels for $Z'_{B-L}$
900032 : onMode = off

Open the decay channel for $Z'_{B-L}$ for pair heavy neutrino productions only otherwise turn off.
900032: onIfAll = 910012 910012

For first daughter of $Z'_{B-L}$ decay (first heavy neutrino) we use these commands:
Turn off all decay channels for $\nu_h$
910012 : onMode = off
         Turn off all decay channels for $\nu_h$
Open the decay channel for $\nu_h$ for W boson and electron only otherwise turn off.
910012 : onIfAll = -24 11
         Turn off all decay channels for W boson
-24 : onMode = off
Open the decay channel for W boson for positron and electron neutrino only otherwise turn off.
-24 : onIfAll = -11 12

For second daughter of $Z'_{B-L}$ decay (second heavy neutrino) we use these commands:
Turn off all decay channels for $\nu_h$
910012 : onMode = off
Open the decay channel for $\nu_h$ for Z boson and electron neutrino only otherwise turns off
910012 : onIfAll = 23 12
Turn off all decay channels for Z boson
23 : onMode = off
Open the decay channel for Z boson to muon and anti-muon
only otherwise turn off.
910012 : onIfAll = 13 -13

After executing these command we get the final state for every event with four charged leptons and missing energy:
$$\mu^- \mu^+ e^- e^+ + \not{E}$$

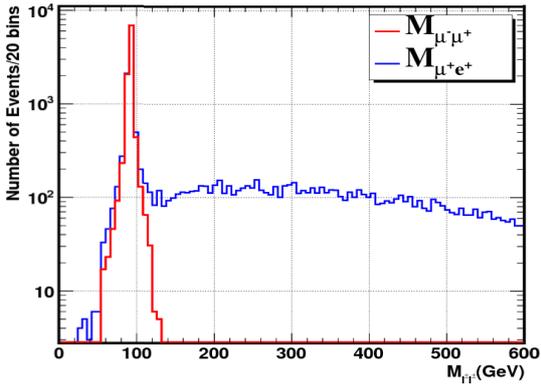

(a)

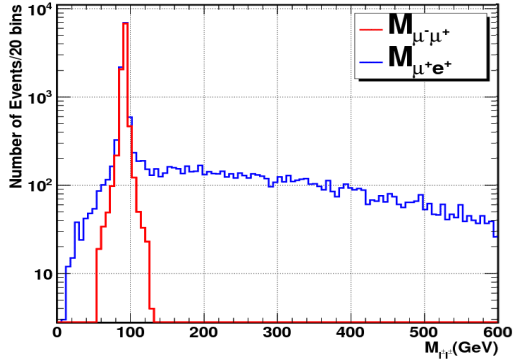

(b)

**FIG. 9.** Distribution n of the invariant mass of the charged lepton pairs in the final state (a) for heavy neutrino mass = 2000 GeV and (b) for heavy neutrino mass = 400 GeV for LHC √s = 14 TeV CM energy, $Z'_{B-L}$ mass = 1.5 TeV and $g'_1$ = 0.2

From fig. (9) We note that there is a clear peak for $M_{\mu^-\mu^+}$ invariant mass for two muons curve where the Z boson mass decay in our simulation process to muon pairs through the Drell Yang mechanism. Z boson production basically comes from decay of one heavy neutrino. The peak values in fig. (9) (a) and (b) are the same for two different heavy neutrino mass 200 GeV and 400 GeV.Hence,from heavy neutrino decay we can reconstruct the mass of Z boson. Another curve is given in fig. 9 a and b which represent the invariant mass of the different leptons $M_{e^+\mu^+}$. The electron comes from the primary decay of the heavy neutrino and the positron come from weak gauge bosons W while the muons come from the weak gauge bosons Z. Fig.(10) represent the transverse momentum for different leptons (electron and muon) produced for the decay of different heavy neutrino masses and we note that the electron has a higher momentum than muon where the electron comes from primary decay of the heavy neutrino while the muon comes from the Z boson decay which basically comes from one of heavy neutrino pair.

The missing $P_T$ distribution in Fig. 11 represents the light neutrinos in the final state where one light neutrino come from the primary decay of the heavy neutrino and another one comes from W boson decay and these two light neutrinos act as missing energy from heavy neutrino pair decay.

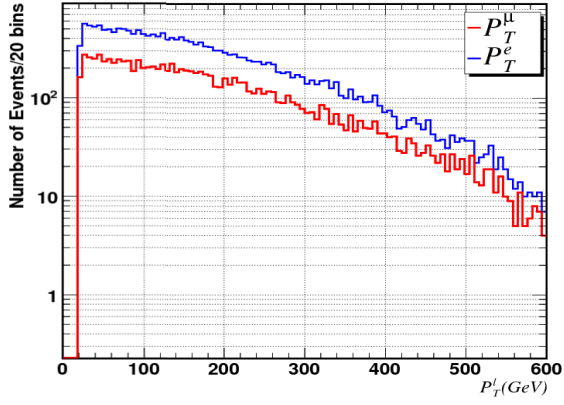

(a)

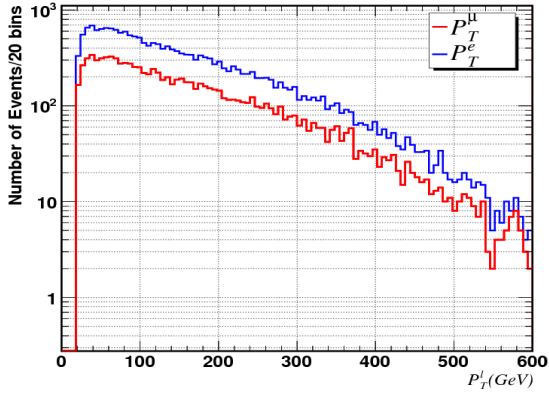

(b)

**FIG. 10.** Distribution of transverse momentum for the different charged leptons in the final state (a) for heavy neutrino mass = 200 GeV and (b) for heavy neutrino mass = 400 GeV for LHC $\sqrt{s}$ = 14 TeV CM energy, $Z'_{B-L}$ mass = 1.5 TeV and $g'_1$ = 0.2 .

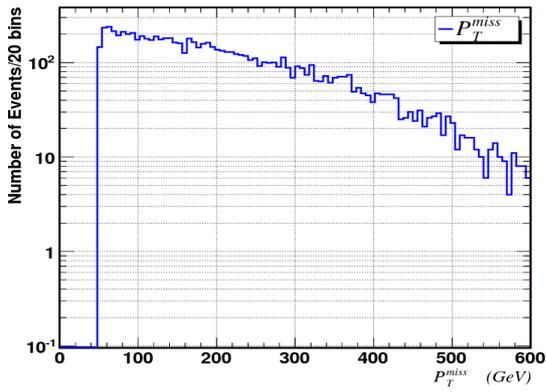

(a)

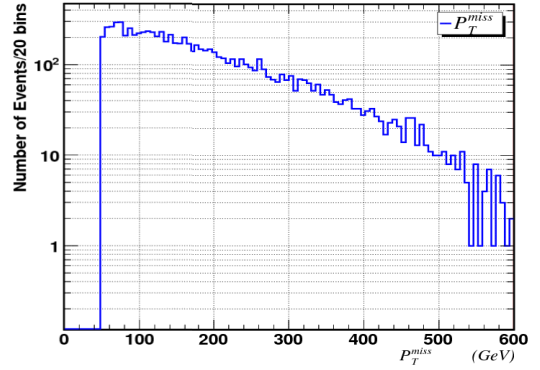

(b)

**FIG. 11.** Distribution of the missing transverse momentum, for the 4l + $\not{E}$ signal : (a) for heavy neutrino mass = 200 GeV and (b) for heavy neutrino mass = 400 GeV for LHC $\sqrt{s}$ = 14 TeV CM energy, $Z'_{B-L}$ ma ass = 1.5 TeV and $g'_1$ = 0.2 .

### III. CONCLUSION

In this work we have presented an analysis For LHC discovery potential for the heavy neutrino in TeV scale using B -L extension of the SM. We analyzed the phenomenology using Monte Carlo simulation data produced from MadGraph5/ Madevent, PYTHIA8 and Calchep and we used ROOT data analysis program to produce different curves for the relationship between specific parameters. From our simulation for this process one can search for heavy neutrinos via a very clean signal for four charged leptons and missing energy at LHC.

### ACKNOWLEDGEMENTS

It is a pleasure to thank T. Sjostrand, for useful discussions of PYTHIA, L. Basso and C. Duhr for useful discussions of B-L model and J. Alwall for useful discussions of MadGraph5/MadEvent.

### REFERENCES

[1] L. Basso, A minimal extension of the Standard Model with B.L gauge symmetry, (Master Thesis,Universit`a degli Studi di Padova, 2007).

[2] J. Collot, A. Ferrari ATL-PHYS-98-124.

[3] S. N. Gninenko June 5, 2007.

[4] K. Huitu, S. Khalil, H. Okada and S. K. Rai, Phys. Rev. Lett. 101 (2008) 181802.

[5] Lorenzo Basso, arXiv:0812.4313v1 [hep-ph] 22 Dec 2008.

[6] L. Basso, (PhD Thesis, university of Southampton).